\documentstyle[aps,preprint,epsf,graphicx]{revtex}
\tighten
\draft

\begin{document}

\def\simge{\hspace*{0.2em}\raisebox{0.5ex}{$>$}
     \hspace{-0.8em}\raisebox{-0.3em}{$\sim$}\hspace*{0.2em}}
\def\simle{\hspace*{0.2em}\raisebox{0.5ex}{$<$}
     \hspace{-0.8em}\raisebox{-0.3em}{$\sim$}\hspace*{0.2em}}
\def\bra#1{{\langle#1\vert}}
\def\ket#1{{\vert#1\rangle}}
\def\coeff#1#2{{\scriptstyle{#1\over #2}}}
\def\undertext#1{{$\underline{\hbox{#1}}$}}
\def\hcal#1{{\hbox{\cal #1}}}
\def\sst#1{{\scriptscriptstyle #1}}
\def\eexp#1{{\hbox{e}^{#1}}}
\def\rbra#1{{\langle #1 \vert\!\vert}}
\def\rket#1{{\vert\!\vert #1\rangle}}
\def\lsim{{ <\atop\sim}}
\def\gsim{{ >\atop\sim}}
\def\nubar{{\bar\nu}}
\def\psibar{{\bar\psi}}
\def\Gmu{{G_\mu}}
\def\alr{{A_\sst{LR}}}
\def\wpv{{W^\sst{PV}}}
\def\evec{{\vec e}}
\def\notq{{\not\! q}}
\def\notk{{\not\! k}}
\def\notp{{\not\! p}}
\def\notpp{{\not\! p'}}
\def\notder{{\not\! \partial}}
\def\notcder{{\not\!\! D}}
\def\notA{{\not\!\! A}}
\def\notv{{\not\!\! v}}
\def\Jem{{J_\mu^{em}}}
\def\Jana{{J_{\mu 5}^{anapole}}}
\def\nue{{\nu_e}}
\def\mn{{m_\sst{N}}}
\def\mns{{m^2_\sst{N}}}
\def\me{{m_e}}
\def\mes{{m^2_e}}
\def\mq{{m_q}}
\def\mqs{{m_q^2}}
\def\mz{{M_\sst{Z}}}
\def\mzs{{M^2_\sst{Z}}}
\def\ubar{{\bar u}}
\def\dbar{{\bar d}}
\def\sbar{{\bar s}}
\def\qbar{{\bar q}}
\def\sstw{{\sin^2\theta_\sst{W}}}
\def\gv{{g_\sst{V}}}
\def\ga{{g_\sst{A}}}
\def\pv{{\vec p}}
\def\pvs{{{\vec p}^{\>2}}}
\def\ppv{{{\vec p}^{\>\prime}}}
\def\ppvs{{{\vec p}^{\>\prime\>2}}}
\def\qv{{\vec q}}
\def\qvs{{{\vec q}^{\>2}}}
\def\xv{{\vec x}}
\def\xpv{{{\vec x}^{\>\prime}}}
\def\yv{{\vec y}}
\def\tauv{{\vec\tau}}
\def\sigv{{\vec\sigma}}
\def\sst#1{{\scriptscriptstyle #1}}
\def\gpnn{{g_{\sst{NN}\pi}}}
\def\grnn{{g_{\sst{NN}\rho}}}
\def\gnnm{{g_\sst{NNM}}}
\def\hnnm{{h_\sst{NNM}}}

\def\xivz{{\xi_\sst{V}^{(0)}}}
\def\xivt{{\xi_\sst{V}^{(3)}}}
\def\xive{{\xi_\sst{V}^{(8)}}}
\def\xiaz{{\xi_\sst{A}^{(0)}}}
\def\xiat{{\xi_\sst{A}^{(3)}}}
\def\xiae{{\xi_\sst{A}^{(8)}}}
\def\xivtez{{\xi_\sst{V}^{T=0}}}
\def\xivteo{{\xi_\sst{V}^{T=1}}}
\def\xiatez{{\xi_\sst{A}^{T=0}}}
\def\xiateo{{\xi_\sst{A}^{T=1}}}
\def\xiva{{\xi_\sst{V,A}}}

\def\rvz{{R_\sst{V}^{(0)}}}
\def\rvt{{R_\sst{V}^{(3)}}}
\def\rve{{R_\sst{V}^{(8)}}}
\def\raz{{R_\sst{A}^{(0)}}}
\def\rat{{R_\sst{A}^{(3)}}}
\def\rae{{R_\sst{A}^{(8)}}}
\def\rvtez{{R_\sst{V}^{T=0}}}
\def\rvteo{{R_\sst{V}^{T=1}}}
\def\ratez{{R_\sst{A}^{T=0}}}
\def\rateo{{R_\sst{A}^{T=1}}}

\def\mro{{m_\rho}}
\def\mks{{m_\sst{K}^2}}
\def\mpi{{m_\pi}}
\def\mpis{{m_\pi^2}}
\def\mom{{m_\omega}}
\def\mphi{{m_\phi}}
\def\Qhat{{\hat Q}}

\def\FOS{{F_1^{(s)}}}
\def\FTS{{F_2^{(s)}}}
\def\GAS{{G_\sst{A}^{(s)}}}
\def\GES{{G_\sst{E}^{(s)}}}
\def\GMS{{G_\sst{M}^{(s)}}}
\def\GATEZ{{G_\sst{A}^{\sst{T}=0}}}
\def\GATEO{{G_\sst{A}^{\sst{T}=1}}}
\def\mdax{{M_\sst{A}}}
\def\mustr{{\mu_s}}
\def\rsstr{{r^2_s}}
\def\rhostr{{\rho_s}}
\def\GEG{{G_\sst{E}^\gamma}}
\def\GEZ{{G_\sst{E}^\sst{Z}}}
\def\GMG{{G_\sst{M}^\gamma}}
\def\GMZ{{G_\sst{M}^\sst{Z}}}
\def\GEn{{G_\sst{E}^n}}
\def\GEp{{G_\sst{E}^p}}
\def\GMn{{G_\sst{M}^n}}
\def\GMp{{G_\sst{M}^p}}
\def\GAp{{G_\sst{A}^p}}
\def\GAn{{G_\sst{A}^n}}
\def\GA{{G_\sst{A}}}
\def\GETEZ{{G_\sst{E}^{\sst{T}=0}}}
\def\GETEO{{G_\sst{E}^{\sst{T}=1}}}
\def\GMTEZ{{G_\sst{M}^{\sst{T}=0}}}
\def\GMTEO{{G_\sst{M}^{\sst{T}=1}}}
\def\lamd{{\lambda_\sst{D}^\sst{V}}}
\def\lamn{{\lambda_n}}
\def\lams{{\lambda_\sst{E}^{(s)}}}
\def\bvz{{\beta_\sst{V}^0}}
\def\bvo{{\beta_\sst{V}^1}}
\def\Gdip{{G_\sst{D}^\sst{V}}}
\def\GdipA{{G_\sst{D}^\sst{A}}}
\def\fks{{F_\sst{K}^{(s)}}}
\def\FIS{{F_i^{(s)}}}
\def\fpi{{F_\pi}}
\def\fk{{F_\sst{K}}}

\def\RAp{{R_\sst{A}^p}}
\def\RAn{{R_\sst{A}^n}}
\def\RVp{{R_\sst{V}^p}}
\def\RVn{{R_\sst{V}^n}}
\def\rva{{R_\sst{V,A}}}
\def\xbb{{x_B}}

\def\PR#1{{{\em   Phys. Rev.} {\bf #1} }}
\def\PRC#1{{{\em   Phys. Rev.} {\bf C#1} }}
\def\PRD#1{{{\em   Phys. Rev.} {\bf D#1} }}
\def\PRL#1{{{\em   Phys. Rev. Lett.} {\bf #1} }}
\def\NPA#1{{{\em   Nucl. Phys.} {\bf A#1} }}
\def\NPB#1{{{\em   Nucl. Phys.} {\bf B#1} }}
\def\AoP#1{{{\em   Ann. of Phys.} {\bf #1} }}
\def\PRp#1{{{\em   Phys. Reports} {\bf #1} }}
\def\PLB#1{{{\em   Phys. Lett.} {\bf B#1} }}
\def\ZPA#1{{{\em   Z. f\"ur Phys.} {\bf A#1} }}
\def\ZPC#1{{{\em   Z. f\"ur Phys.} {\bf C#1} }}
\def\etal{{{\em   et al.}}}

\def\delalr{{{delta\alr\over\alr}}}
\def\pbar{{\bar{p}}}
\def\lamchi{{\Lambda_\chi}}
\newcommand{\amulbl}{a_\mu^{\sst{LL}}}

\preprint{
\noindent
\hfill
\begin{minipage}[t]{3in}
\begin{flushright}
KRL-MAP-283\\
\vspace*{.7in}
\end{flushright}
\end{minipage}
}

\title{What Do We Know About the Strange Magnetic Radius?}

\author{H.-W. Hammer$^{a}$, S.J. Puglia$^{a}$, M.J. Ramsey-Musolf$^{b,c}$,
and Shi-Lin Zhu$^{b}$
\\[0.3cm]
}
\address{
$^a$ Department of Physics, The Ohio State University, Columbus,
OH 43210\ USA \\
$^b$ California Institute of Technology,
Pasadena, CA 91125\ USA\\
$^c$ Department of Physics, University of Connecticut, Storrs, CT 06269\ USA
}


\maketitle

\begin{abstract}

We analyze the $q^2$-dependence of the strange magnetic form factor,
$\GMS(q^2)$, using heavy baryon
chiral perturbation theory (HB$\chi$PT) and dispersion relations. We find
that in HB$\chi$PT a significant
cancellation occurs between the ${\cal O}(p^2)$ and ${\cal O}(p^3)$ loop
contributions. Consequently, the
slope of $\GMS$ at the origin displays an enhanced sensitivity to an
unknown ${\cal O}(p^3)$ low-energy
constant. Using dispersion theory, we estimate the magnitude of this
constant, show that it may have a
natural size, and conclude that the low-$q^2$ behavior of $\GMS$ could be
dominated by nonperturbative
physics. We also discuss the implications for the interpretation of
parity-violating electron scattering
measurements used to measure $\GMS(q^2)$.

\end{abstract}



\vspace{0.3cm}

\pagenumbering{arabic}

The flavor structure of  low-energy nucleon properties ({\em e.g.}, mass,
spin, {\em etc.})
continues to be a subject of considerable interest in hadron physics. In
particular, evidence exists that
strange quarks may play a non-negligible role in such low-energy
properties\cite{squarks}. The strange quark
contribution to the nucleon's electromagnetic structure is under
investigation using parity-violating
electron scattering experiments at MIT-Bates \cite{sample}, the Jefferson
Lab (JLab) \cite{happex,g0}, and
Mainz \cite{mainz}. These experiments have produced the first
determinations of the strange magnetic form
factor,
$\GMS(q^2)$ \cite{sample}, as well as a linear combination of $\GMS(q^2)$
and the strange electric form
factor, $\GES(q^2)$ \cite{happex}.

Theoretical interest has focused largely on the leading moments of these
form factors: the strange magnetic
moment,
$\mu_s=\GMS(0)$ and the corresponding electric and magnetic radii, $\langle
r^2_s \rangle_{E,M} = 6 d
G_\sst{E,M}^{(s)}(q^2)/dq^2$ at $q^2=0$ (note that we do not normalize the
radii to the
$q^2=0$ values of the form factors). While the
radii are intrinsically interesting, knowledge of
$\langle r^2_s \rangle_{M}$ is also needed in order to extract 
$\mu_s$ from experiment, since $\GMS$ can be determined
experimentally only for non-zero $q^2$. In order to extrapolate $\GMS$ to
the photon point, one might hope
to rely on chiral perturbation theory, which is in principle well-suited to
characterizing the leading
$q^2$-dependence of form factors. While $\langle r^2_s \rangle_{M}$ 
is nominally of chiral order $p^3$ (we count the
magnetic moment operator as being ${\cal O}(p)$), the leading order
contribution
from kaon loops is proportional to
$1/m_K\sim 1/\sqrt{m_s}$ and, thus, of ${\cal O}(p^2)$. At this order,
there exist no analytic (in
quark mass) operators and, consequently, no counterterm or corresponding
low-energy constant. Thus,
as observed by the authors of Ref. \cite{meissner}, one is able to make a
parameter-free prediction for $\langle
r^2_s \rangle_{M}$ at this order which may be 
used to perform a model-independent extrapolation of $\GMS$ to the
photon point. This procedure has been used to extract a value of
$\mu_s=0.01\pm 0.29\pm 0.31\pm 0.07$ from the value of the form factor
obtained by the SAMPLE Collaboration at $q^2= -0.1$
(GeV/$c$)$^2$: $\GMS(q^2=-0.1) = 0.14\pm 0.29\pm 0.31$ \cite{sample} (The
last error in $\mu_s$
corresponds to the theoretical extrapolation uncertainty quoted in Ref.
\cite{meissner}.)

Here, we compute the ${\cal O}(p^3)$  loop contributions to $\GMS(q^2)$ in
heavy baryon chiral perturbation
theory (HB$\chi$PT). We show that the ${\cal O}(p^3)$
contributions to $\langle r^2_s \rangle_{M}$ largely
cancel the ${\cal O}(p^2)$ term. In addition, we observe that a new
magnetic radius operator arises at this
order, whose coefficient cannot presently be determined apart from a
determination of $\GMS$ itself.
Because of the cancellation between the ${\cal O}(p^2)$ and ${\cal O}(p^3)$
terms, the relative importance
of this coefficient, or low-energy constant (LEC), is enhanced over what
one might otherwise expect.
Using a dispersion theory analysis of
$\GMS$, we estimate this LEC and argue that it could give the dominant
contribution to $\langle r^2_s \rangle_{M}$. As a
corollary, we suggest that the low-momentum behavior of $\GMS$ is governed
by non-perturbative physics and
observe that  one cannot presently extract $\mu_s$ from experiment in a
model-independent manner.

In HB$\chi$PT, the contributions to $\GMS(q^2)$ through ${\cal O}(p^3)$
are generated by the graphs of Fig.~1. The ${\cal O}(p)$ contributions to
$\mu_s$ arise from insertion of
the lowest-order magnetic operator in Fig.~1(a). The ${\cal O}(p^2)$
contributions nominally arise from Figs.~1(b-e). 
However, graphs 1(c-e) contribute only to the charge form factor, so
that the ${\cal O}(p^2)$
contributions to $\mu_s$ arise entirely from Fig.~1(b). The ${\cal O}(p^3)$
contributions to $\mu_s$ arise
from several sources: insertion of the tree-level magnetic moment operator
in Figs.~1(f,g); operators
containing two-derivatives (Fig.~1(h)); $1/M_N$, or \lq\lq  recoil",
corrections in Figs.~1(i-k) (denoted by the
\lq\lq $\times$"); and magnetic operators  proportional to $m_s$ (Fig.~1(l)).
We note that the sum of Figs.~1(j) and 1(k) vanishes for $\GMS$.

Similarly, the leading contribution to the slope of $\GMS(q^2)$ at the
origin also arises from
Fig.~1(b) and is ${\cal O}(p^2)$. The ${\cal O}(p^3)$ contributions to the
slope are given by the
sum of Fig.~1(h) and 1(i), along with the insertion of the magnetic radius
operator in Fig.~1(a). One must
also include the dependence of Fig.~1(b) on $v\cdot q=q^2/M_N$.

The formalism for evaluating these diagrams in HB$\chi$PT has been
discussed extensively elsewhere, so we do
not give all the relevant formulae here (see, {\em e.g.}, Ref.
\cite{puglia} and references therein).
However, in order to define our normalization for the LEC's we give
expressions for the relevant magnetic
moment and radius operators:
\begin{eqnarray}
\label{eq:lmm}
{\cal L}_1 & = & {eb_0\over\lamchi}\epsilon_{\mu\nu\rho\sigma} v^\rho {\rm
Tr}\left({\bar B} S^\sigma
B\right) Z^{\mu\nu} \\
\nonumber
&&+{e\over\lamchi}\epsilon_{\mu\nu\rho\sigma} v^\rho\Bigl\{ b_+{\rm
Tr}\left({\bar B} S^\sigma
\{Q,B\}\right) +b_-{\rm Tr}\left({\bar B} S^\sigma
[Q,B]\right)\Bigr\} F^{\mu\nu}\\
\nonumber
{\cal L}_{SB} & = & {e\over\lamchi}\epsilon_{\mu\nu\rho\sigma} v^\rho
F^{\mu\nu} \Bigl\{
b_3{\rm Tr}\left({\bar B} S^\sigma [[Q,B],{\cal M}]\right) \\
\label{eq:lsb}
&&+b_4{\rm Tr}\left({\bar B} S^\sigma \{[Q,B],{\cal M}\}\right)
+b_5{\rm Tr}\left({\bar B} S^\sigma [\{Q,B\},{\cal M}]\right)\\
\nonumber
&&+b_6{\rm Tr}\left({\bar B} S^\sigma \{\{Q,B\},{\cal M}\}\right)
+b_7{\rm Tr}\left({\bar B} S^\sigma B\right){\rm Tr}\left({\cal
M}Q\right)\Bigr\}\\
\nonumber
&&+{e b_8\over\lamchi}\epsilon_{\mu\nu\rho\sigma} v^\rho Z^{\mu\nu}
{\rm Tr}\left({\bar B} S^\sigma B\right)\ {\rm Tr}\left({\cal M}\right)\\
\label{eq:lmr}
{\cal L}_2 & = & {eb_0^r\over\Lambda_\chi^3}\epsilon_{\mu\nu\rho\sigma}
v^\rho{\rm Tr}\left({\bar B}
S^\sigma B\right) {\partial}^2 Z^{\mu\nu}\\
\nonumber
&&+{e\over\lamchi}\epsilon_{\mu\nu\rho\sigma} v^\rho\Bigl\{ b_+^r{\rm
Tr}\left({\bar B} S^\sigma
\{Q,B\}\right) +b_-^r{\rm Tr}\left({\bar B} S^\sigma
[Q,B]\right) \Bigr\} {\partial}^2 F^{\mu\nu}
\end{eqnarray}
where $B$ denotes the baryon field octet, $Q= {\rm diag}(2/3, -1/3, -1/3)$
(at lowest order), ${\cal M}={\rm
diag}(0,0,1)$ is the SU(3) symmetry-breaking matrix associated with the
light quark masses, $v^\rho$
and $S^\sigma$ are the baryon velocity and spin, respectively, $F^{\mu\nu}$
is the photon field
strength tensor, and $Z^{\mu\nu}$ is the corresponding tensor for a source
coupling to baryon number
current\footnote{A typo appears in Eq.~(17) of Ref. \cite{puglia}. The term
proportional to $b_6$ should
contain a double anti-commutator.}. The chiral scale
$\lamchi\equiv 4\pi F_\pi$, with $F_\pi = 92.4$ MeV. The full chiral
structure of the electromagnetic
charge operator is obtained by making the replacement
$Q\to\frac{1}{2}(\xi^{\dag}Q\xi + \xi Q\xi^{\dag})$,
where $\xi =
\exp(i{\tilde
\Pi}/F_\pi)$ and ${\tilde\Pi}$ is the octet of pseudscalar mesons. Note
that the symmetry-breaking
operators in ${\cal L}_{SB}$ are suppressed by two powers of $p$ relative
to ${\cal L}_1$. Following
standard conventions, however, we absorb this suppression into the
definition of the constants
$b_{3-8}$. Indeed, when $b_{\pm}$ and the octet constants $b_{3-7}$ are
determined from fits to octet baryon
magnetic moments\cite{puglia}, the latter are generally suppressed by an
order of magnitude relative to the
former, as one would expect from the $p^2$ suppression of ${\cal L}_{SB}$.

The two-derivative operators
have the structure
\begin{eqnarray}
\label{eq:twoderiv}
{\cal L}_{MB} & = & {4i\over\lamchi}\epsilon_{\mu\nu\rho\sigma}
v^\rho\Bigl\{ b_9{\rm Tr}\left({\bar B}
S^\sigma A^\mu\right) {\rm Tr}\left(A^\nu B\right) \\
\nonumber
&& + b_{10}{\rm Tr}\left({\bar B} S^\sigma[A^\mu,A^\nu] B\right) +
b_{11}{\rm Tr}\left({\bar B} S^\sigma\{A^\mu,A^\nu\} B\right)\Bigr\}\ \ \ \ ,
\end{eqnarray}
where $A^\mu = (i/2)(\xi^{\dag}\partial^\mu \xi - \xi\partial^\mu \xi^{\dag})$.

Using these normalizations, and expanding
\begin{equation}
\GMS(q^2) = \mu_s +\frac{1}{6} \langle r^2_s \rangle_{M}\; q^2 +\cdots\ \ \ ,
\end{equation}
we obtain
\begin{eqnarray}
\label{eq:mus1}
\mu_s & = & \left({2 M_N\over\lamchi}\right)\Biggl\{ b_s  \left( 1 +
\sum_{X=\pi,K,\eta}(\gamma^{(X)}
-\lambda^{(X)})\frac{m_X^2}{\Lambda_\chi^2}\ln\frac{m_X^2}{\mu^2}\right)\\
\nonumber
&&+(5D^2-6DF+9F^2)\left(\frac{\pi}{3}\frac{m_K}{\lamchi} -
\frac{5}{6}\frac{m_K^2}{M_N\lamchi}\ln\frac{m_K^2}{\mu^2}\right)\\
\nonumber
&&+b_8-2(b_3+b_4-\frac{1}{3}b_5-\frac{1}{3}b_6-b_7)-h
\frac{m_K^2}{\Lambda_\chi^2}\ln\frac{m_K^2}{\mu^2}
\Biggr\}\\
\nonumber
\\
\label{eq:rsm1}
\langle r^2_s\rangle _{M} & = & -\frac{6}{\Lambda_\chi^2}\Biggl\{\left({2
M_N\over\lamchi}\right)b_s^r
+\frac{1}{18}(5D^2-6DF+9F^2)\left(\frac{\pi M_N}{m_K} +
7\ln\frac{m_K}{\mu}\right)\\
\nonumber
&&-\left({M_N\over 9\lamchi}\right)h\ln\frac{m_K}{\mu}
\Biggr\}\ \ \ \,
\end{eqnarray}
where
\begin{eqnarray}
\nonumber
b_s & = & b_0 - 2[b_- - (b_+/3)] \\
\label{eq:bsr}
b_s^r & = & b_0^r - 2[b_-^r -(b_+^r/3)]\\
\nonumber
h & = & b_9-2b_{10}+6b_{11}
\end{eqnarray}
and where $\mu$ is the subtraction scale associated with the divergence of
the Feynman diagrams, $D$ and
$F$ are the SU(3) axial current reduced matrix elements, and $\gamma^{(X)}$
and $\lambda^{(X)}$ are
known functions of $D$ and $F$ \cite{puglia}. The dependence of $b_s$
($b_s^r$) on the SU(3) singlet
constant $b_0$ ($b_0^r$) and octet constants $b_{\pm}$ ($b_{\pm}^r$), as
well as the
presence of the $b_{3-8}$ terms in $\mu_s$ arise
from the group structure of the strange quark vector current:
\begin{equation}
\label{eq:vec}
{\bar s}\gamma_\mu s = J_\mu^B -2 J_\mu^{I=0}(EM)\ \ \ ,
\end{equation}
where $J_\mu^B$ is the SU(3) singlet, or baryon number, current and
$J_\mu^{I=0}(EM)$ is the
isoscalar electromagnetic (EM) current, an SU(3) octet operator. Nucleon
matrix elements of the latter
receive contributions from the $b_{3-7}$ operators in Eq.~(\ref{eq:lsb}),
as evaluated in
Ref. \cite{puglia}. The operators proportional to $b_{0,8}$ and $b_0^r$
contribute to $\bra{p}
J_\mu^B\ket{p}$. While values for the octet LEC's $b_{\pm}$, $b_{\pm}^r$,
and $b_{3-7}$ can be obtained
from measured octet baryon magnetic moments, one has no independent
determination of the singlet LEC's
$b_0$, $b_0^r$, and $b_8$ (apart from measurements of $\GMS$ itself).
Note also that the LEC's are functions of the subtraction scale,
$b_i=b_i(\mu)$ and
$b_s=b_s(\mu)$.

The contributions through ${\cal O}(p^2)$
for both $\mu_s$ and $\langle r^2_s \rangle_{M}$ have been computed in Refs.
\cite{meissner,ito}. In the case of $\langle
r^2_s \rangle_{M}$, the ${\cal O}(p^2)$ contribution contains the factor
$M_N/m_K$. The contributions of ${\cal
O}(p^3)$ have not been computed previously. As a check on our results, we
compare with the integrals appearing in the relativistic
calculation of the isovector electromagnetic form factors in Ref.
\cite{kubis}. After replacing $m_\pi$ by $m_K$, we find agreement
with the non-analytic contributions through ${\cal O}(p^3)$. Note
that symmetry-breaking effects which
break the degeneracy of the pseudoscalar decay constants ($F_\pi\not= F_K
\not= F_\eta$) contribute
at higher order than we consider here, so we use only $F_\pi$ in our
numerical evaluation below. Similarly,
the lowest-order effect of the $M_{\Lambda,\Sigma}$ and $M_N$
non-degeneracy is analytic in $m_s$ and can
be absorbed entirely in the LEC's $b_0$, $b_\pm$, $b_0^s$, {\em etc.}.
Finally, as in Ref. \cite{meissner},
we have included only octet baryon intermediate states. While inclusion of
decuplet states will likely
affect the precise numerical values of the ${\cal O}(p^2)$ and ${\cal
O}(p^3)$ contributions, we do not
expect modification of our qualitative conclusions.

The low-$|q^2|$ behavior of $\GMS$
depends on several LEC's, including $b_s$, $b_s^r$, and $b_{8-11}$.
To determine the relative importance of the LEC's and loops, we take
$D=0.75$ and $F=0.5$,  set
$\mu = 1$ GeV, and derive values for $b_3\ldots b_7$ from the octet
magnetic moments.
In doing so, we evaluate the combination of constants $h$ as in Ref.
\cite{stein}.
The result gives
\begin{eqnarray}
\label{eq:mus2}
\mu_s & = & 2.5 \left[ b_s(\mu = 1\ {\rm GeV}) + 0.6 b_8(\mu = 1\ {\rm
GeV})\right]+ 1.2 \\
\label{eq:rsm2}
\langle r^2_s\rangle _{M} & = & -\left[ 0.04 + 0.3 b_s^r(\mu = 1\ {\rm
GeV}) \right] \ \ {\rm fm}^2 \ \ \ ,
\end{eqnarray}
where the LEC-independent contributions $1.2$ (in $\mu_s$) and $-0.04$
fm$^2$ (in $\langle r^2_s \rangle_{M}$) arise
entirely from the loop graphs and symmetry-breaking LEC's. We note in
particular the
relative sensitivity of $\langle r^2_s \rangle_{M}$
to loops and the LEC $b_s^r$. Were the latter to have a \lq\lq natural"
size, $|b_s^r|\sim 1$,
its contribution would dominate the slope of $\GMS$ and the origin. Indeed,
an uncertainty in
this LEC of $\Delta b_s^r=\pm 1$ corresponds to a $\pm 0.3 $ fm$^2$
uncertainty in the radius.
This situation contrasts starkly with the result at 
${\cal O}(p^2)$, which gives $\langle r^2_s\rangle_{M} = -0.16$
fm$^2$ (arising entirely from Fig.~1(b)). In
short, inclusion of the ${\cal O}(p^3)$ loop contribution produces a
substantial cancellation with the
${\cal O}(p^2)$ term, thereby enhancing the relative sensitivity of
$\langle r^2_s\rangle _{M}$ to the LEC
$b_s^r$. While one generally expects a suppression of higher-order effects,
the rate of convergence
for the strangeness form factors is governed by $m_K/\lamchi\sim 1/2$ and,
thus, is rather slow.
Consequently, the occurance of significant cancellations between successive
orders as we find here is
not entirely surprising.

At this time, it is not possible to determine $\mu_s$ and 
$\langle r^2_s \rangle_{M}$ from existing measurements.
As evident from Eq.~(\ref{eq:vec}) the strange quark vector current
contains both SU(3)-octet and
SU(3)-singlet components. Contributions to
to $b_s+0.6 b_8$ and $b_s^r$ from the octet component are known, while the
singlet component is
undetermined. In principle, one might attempt to determine $\mu_s$ -- and,
thus, $ b_s + 0.6 b_8$ --
from the published value of
$\GMS(q^2=-0.1)$, using the lowest-order HB$\chi$PT result for $\langle
r^2_s \rangle_{M}$ as was done in Refs.
\cite{sample,meissner}. However, as the results in 
Eqs.~(\ref{eq:rsm1},\ref{eq:rsm2}) indicate, such a
procedure could be misleading. Given the sensitivity of $\langle 
r^2_s \rangle_{M}$ to $b_s^r$, neither the magnitude
nor the sign of the magnetic radius can be determined at present in a
model-independent manner. Thus, the correct extrapolation of
$\GMS$ from $q^2=-0.1 $ (GeV/$c)^2$ to the photon point is simply not known
at present.

To illustrate this situation, we plot in Fig.~2 $\GMS$ as a function of
$Q^2=-q^2$, showing the published
result at $Q^2= 0.1$ (GeV/$c)^2$. We also show various possible
extrapolations to $Q^2=0$. The dashed lines
indicate the extrapolation for $b_s^r=-1$ and the dot-dashed lines give the
extrapolation for $b_s^r=+1$.
Taking these two possibilities as reasonable, though not rigorous,
extremes, one would infer that the
strange magnetic moment falls in the range $0.7 \gtrsim \mu_s \gtrsim
-0.4$. Using only the leading order
HB$\chi$PT results one would infer $ 0.5 \gtrsim \mu_s \gtrsim -0.3 $. We
emphasize,
however, that larger values for $|b_s^r|$ than used in Fig.~2 would not be
unreasonable, so we would not
want to use $\Delta b_s^r =\pm 1$ as an estimate of the theoretical
uncertainty. For example, changing
$b_s^r$ from
$\pm 1$ to $\pm 2$ increases the
range in $\mu_s$ by $\sim \pm 0.1$. We also note that, while the present
uncertainty in $\mu_s$ is likely dominated by the experimental error, the
relative importance of the $|b_s^r|$-induced uncertainty could become
comparable to the experimental error with more precise data for
$\GMS(q^2=-0.1)$.

In general, LEC's such as $b_s$ and $b_s^r$ characterize the effects of
short-distance physics arising
from scales $r < 1/\lamchi$. Such non-perturbative  effects can, in
principle, be computed using
lattice QCD or dispersion relations. While  no definitive calculation has
been performed using either
approach, it is nevertheless instructive  to compare the implications for
the LEC's from existing analyses.
In the quenched lattice calculation of Ref. \cite{liu}, $\GMS(q^2)$ was
evaluated at five different
kinematic points, starting at $Q^2=0$ and increasing by increments of
roughly $\Delta Q^2 =0.4$
(GeV/$c)^2$. The results at $Q^2=0$ give $\mu_s = -0.36 \pm 0.20$,
corresponding to $b_s+0.6 b_8\approx -0.6
\pm 0.1$. Given both the spacing of the lattice data as well as the size of
the error bars, there exists
considerable lattitude for the shape of $\GMS$ at low-$Q^2$. For example,
the authors of Ref. \cite{liu} fit
the lattice data to a dipole form factor, yielding a radius parameter
$b_s^r(\mu = 1\ {\rm GeV}) =0.23 \pm
0.13$. However, a different shape for $\GMS$ at low-$Q^2$ -- including an
opposite sign slope at the
origin -- would not be inconsistent with the published lattice results.

In the case of dispersion relations, the strange magnetic moment and
magnetic radius are related to
integrals over the imaginary parts of the relevant strange form factors:
\begin{eqnarray}
\label{eq:disp1}
\mu_s & = & \frac{1}{\pi}\ \int_{9 m_\pi^2}^\infty\ \frac{{\rm Im}
F_2^{(s)}(t)}{t}\ dt \\
\nonumber
\\
\label{eq:disp2}
\langle r^2_s\rangle _{M} & = & \frac{6}{\pi}\ \int_{9 m_\pi^2}^\infty\
\frac{{\rm Im} \GMS(t)}{t^2}\ dt\ \
\ .
\end{eqnarray}
Note that since the Sachs form factor $\GMS(q^2)$ is the sum of Dirac and
Pauli form factors,
$\FOS$ and $\FTS$, respectively, and since $\FOS(0)=0$, only the integral
over $\FTS$ contributes to
$\mu_s$.

In each case, the integrands receive contributions from a tower of hadronic
states:
$3\pi$, $5\pi$, $7\pi$, $K{\bar K}$, $\ldots$. In previous
work\cite{KKcont,hammer}, we evaluated the $K{\bar K}$
contribution  using $KN$ scattering and $e^+e^-\to K{\bar K}$ data.  We
find that the
\lq\lq kaon cloud" contribution to both the strange vector form factors and
EM isoscalar
form factors is dominated
by the $\phi(1020)$-resonance. For the purely pionic intermediate states,
$e^+e^-$ data suggest that only
the $3\pi$ channel gives an important contribution and that it is dominated
by the $\omega$-resonance.
Using the known flavor content of the $\omega$, one can follow Ref.
\cite{jaffe} and evaluate its
contribution to $\GMS$.

Inclusion of higher-mass states ({\em e.g.}, $KK\pi$) is necessary in order to
obtain a physically reasonable large-$Q^2$ behavior of vector form factors.
In the case of the
isoscalar magnetic form factor, the sum of the $3\pi$ and $K{\bar K}$
contributions -- along with a single
higher mass pole to approximate the effect of the remaining higher-mass
states -- gives an excellent
description over a wide range for $Q^2$. Since we do not compute the
higher-mass contributions
from first principles, however, we are not confident in evaluating their
contribution to $\GMS$.
Consequently, we quote only the \lq\lq low-mass" ($3\pi$ and $K{\bar K}$)
contributions to the moments in
Eqs.~(\ref{eq:disp1},
\ref{eq:disp2}): $\mu_s=-0.28$ and $\langle r^2_s \rangle_{M}=-0.34$ fm$^2$
\cite{hammer}. These values correspond to
$b_s+0.6 b_8\approx -0.6$ and $b_s^r\approx -1.1$, respectively. A
comparison of these values with the
lattice/dipole fit values is given in Table I. The corresponding low-mass
dispersion
relation prediction for the low-$Q^2$ behavior of $\GMS$ is also shown in
Fig.~2. A reasonable, though not
rigorous, estimate for the uncertainty associated with presently unknown
higher-mass contributions is also
shown (shaded band)\footnote{Additional potential sources of uncertainty
include
those associated with the $\omega$-pole approximation for the $3\pi$
contribution, the analytic continuation used to obtain the spectral functions,
{\em etc.}.}.
For example, a negative slope but with smaller magnitude arises in the pole
approximation analyses of form factors given in Refs.
\cite{jaffe,hammer2}.

We observe that the lattice and \lq\lq low-mass"
dispersion relation determinations of $b_s$ are comparable, while the
dipole fit to lattice data gives
a significantly different prediction for the radius parameter $b_s^r$ than
obtained with the dispersion
relation analysis. We speculate that this difference could be  an artifact
choosing a dipole form
factor to parameterize the lattice evaluation of $\GMS$, a result of the
omission of higher-mass
contributions to the dispersion integral, or both. More importantly, each
of these approaches, which
incorporate both chiral and nonperturbative effects, produce values for the
LEC $b_s^r$ of natural
size, though having opposite sign. In either case, one would conclude that
non-perturbative dynamics
play an important role in the physics of $\GMS$. Ideally, future
theoretical progress will ultimately lead
to consistency between the lattice and dispersion relation treatments of
these non-perturbative effects. Such progress will undoubtedly require
incorporation of the correct chiral structure of $\GMS$ at low-$|q^2|$.
From the experimental side, additional data for $\GMS$ at larger $|q^2|$
may not be sufficient to determine $\mu_s$ and $\langle r_s^2 \rangle_M$, 
since the chiral expansion is valid for a limited $|q^2|$-range 
({\em e.g}, out
to $\sim m_K^2$) and since new operators contribute as one moves successively
further from the photon point. On the other hand, additional low-$|q^2|$ 
data [{\em e.g.}, $|q^2| < 0.05$ (GeV$/c)^2$] from parity-violating
electron scattering experiments could provide an unambiguous determination
of the behavior of strange magnetism at small momentum-transfer.

This work was supported in part under U.S. Department of
Energy contract \#DE-FGO2-00ER41146 and by
the National Science Foundation under award PHY00-71856
and grant No. PHY-0098645.
We thank B.R. Holstein, B. Kubis, R.D. McKeown, U.-G. Mei{\ss}ner,
and T.M. Ito for useful discussions.

\begin{figure}
\begin{center}
\includegraphics[height=3.2in,angle=0,clip=true]{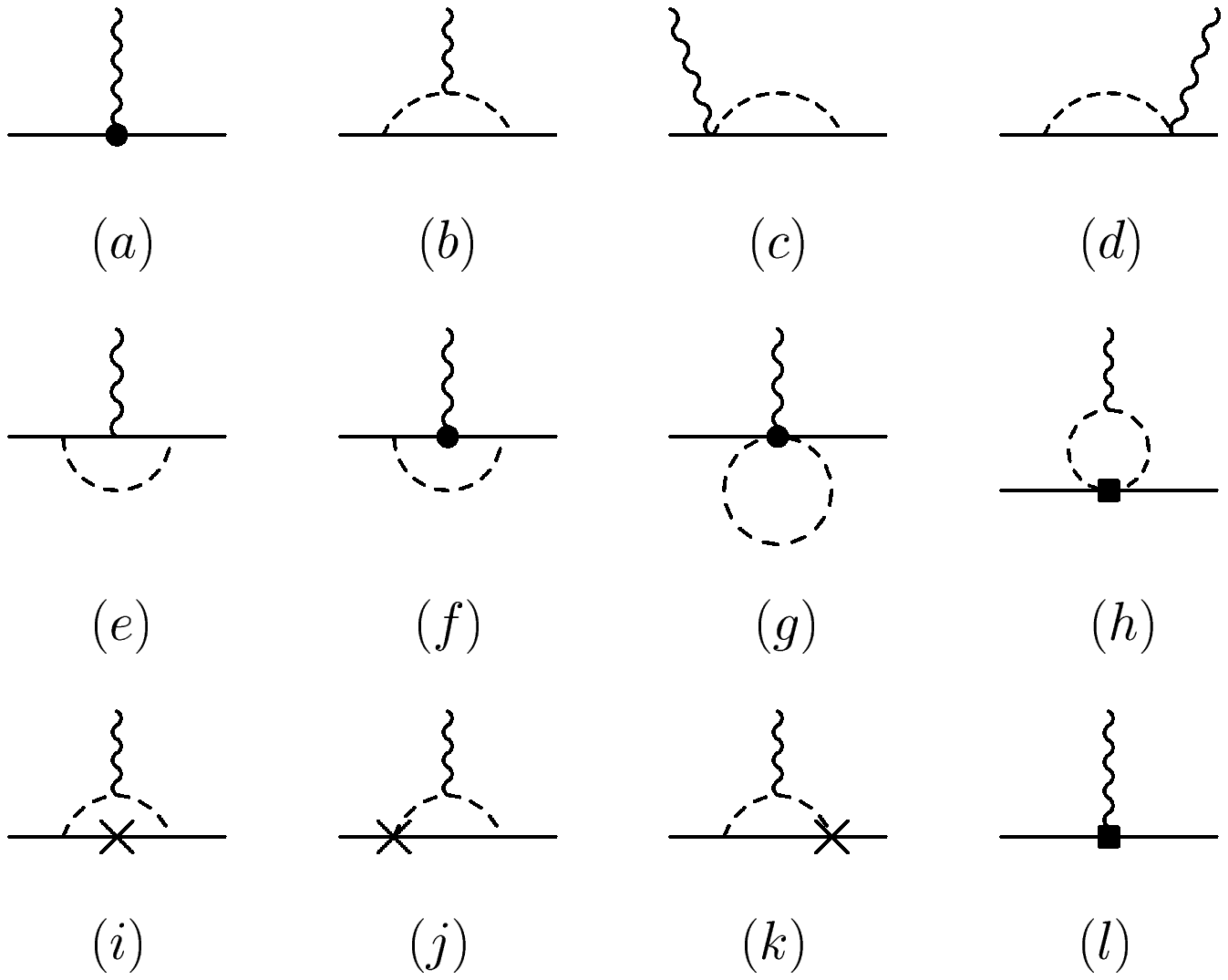}
\end{center}
\caption{\label{Fig1} Contributions to strange magnetic moment and radius.
Circle denotes
leading-order magnetic moment and radius operators, while square indicates
operators
proportional to quark mass or containing two derivatives. The $\times$
indicates a $1/M_N$
insertion. Solid line and dashed lines indicate octet baryons and
pseudoscalar mesons, respectively.}
\end{figure}

\begin{figure}
\begin{center}
\includegraphics[height=3.2in,angle=0,clip=true]{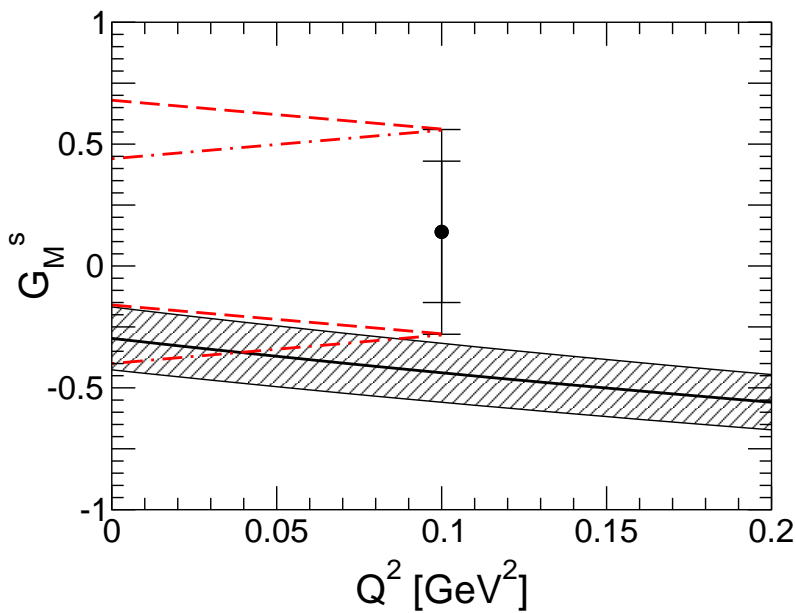}
\end{center}
\caption{\label{Fig2} The $Q^2$-dependence of $\GMS$. Circle at $Q^2=0.1$
(GeV/$c)^2$ indicates the
result from the SAMPLE measurement \protect\cite{sample}.
Inner error bar gives statistical error while outer
error bar combines statistical and systematic errors in quadrature. Dashed
(dot-dashed) lines indicate
extrapolation to $Q^2=0$ using radius parameter $b_s^r=-1$ ($+1$). Solid
line gives \lq\lq low-mass"
dispersion relation result \protect\cite{hammer},
while shaded region indicates possible effects of higher
mass contributions to the dispersion relation.}
\end{figure}

\begin{table}
\begin{center}~
\begin{tabular}{|c|c|c|}\hline
 Method & $b_s+0.6b_8$ & $b_s^r$
\\ \hline
Lattice/Dipole Fit & $-0.6\pm 0.1$  & $0.23\pm 0.13$  \\ \hline
Low-mass Dispersion Relation & $-0.6$  & $-1.1$  \\ \hline
\end{tabular}
\end{center}
\caption{\label{tab1} Low-energy constants for strange magnetic moment
($b_s$) and
strange magnetic radius ($b_s^r$) obtained from (a) lattice QCD calculation of
Ref.~\protect\cite{liu} and (b) $3\pi$ and $K{\bar K}$ contributions using
dispersion relations \protect\cite{hammer}.
The the lattice value for $b_s^r$ was obtained by fitting lattice data to a
dipole form
for $\GMS(q^2)$.}
\end{table}


\begin{thebibliography}{99}

\bibitem{squarks} See, {\em e.g.}, B.W. Filippone and X. Ji, Adv. in Nucl.
Phys. {\bf 26}, 1 (2001)
and J. Gasser, H. Leutwyler, and M.E. Sainio, Phys. Lett. B {\bf 253}, 163
(1991).

\bibitem{sample} R. Hasty, {\em et al.}, the SAMPLE Collaboration, Science
{\bf 290}, 2117 (2000).

\bibitem{happex} K. Aniol {\em et al.}, the HAPPEX Collaboration, Phys.
Lett. B {\bf 509}, 211 (2001).

\bibitem{g0} Jefferson Laboratory experiment 00-006, D. Beck, spokesperson.

\bibitem{mainz} Mainz experiment PVA4, D. von Harrach, spokesperson; F.
Maas, contact person.

\bibitem{meissner} T.R. Hemmert, U.-G. Mei{\ss}ner, and S. Steininger, Phys.
Lett. B {\bf 437}, 184 (1998);
T.R. Hemmert, B. Kubis, and U.-G. Mei{\ss}ner, Phys. Rev. C {\bf 60}, 045501
(1999).

\bibitem{puglia} S.J. Puglia and M.J. Ramsey-Musolf, Phys. Rev. D {\bf 62},
034010 (2000).

\bibitem{ito} M.J. Ramsey-Musolf and H. Ito, Phys. Rev. C {\bf 55}, 3066
(1997).

\bibitem{kubis} B. Kubis and U.-G. Mei{\ss}ner, Nucl. Phys. A {\bf 679}, 698
(2001).

\bibitem{stein} U.-G. Mei{\ss}ner and S. Steininger, Phys. Phys. B {\bf 499},
349 (1997).

\bibitem{liu}S.J. Dong, K.F. Liu, and A.G. Williams, Phys. Rev. D {\bf 58},
074504 (1998).

\bibitem{KKcont} M.J. Ramsey-Musolf and H.-W. Hammer, Phys. Rev. Lett.
 {\bf 80}, 2539 (1998); H.-W. Hammer and M.J. Ramsey-Musolf,
 Phys. Rev. C {\bf 60}, 045205 (1999).

\bibitem{hammer} H.-W. Hammer and M.J. Ramsey-Musolf, Phys. Rev. C {\bf
60}, 045204 (1999).

\bibitem{jaffe}R. L. Jaffe, Phys. Lett. B {\bf 229}, 275 (1989).

\bibitem{hammer2} H.-W. Hammer, U.-G. Mei{\ss}ner, and D. Drechsel, Phys.
Lett. B {\bf 367}, 323 (1996).


\end{thebibliography}
\end{document}